\pdfoutput = 1
\documentclass[twocolumn,showpacs,prb,amsmath,amssymb,floatfix]{revtex4-1}
\usepackage{amsmath,amssymb}
\usepackage{bm}
\usepackage{graphicx}
\usepackage{psfrag}
\usepackage{color}

\begin{document}

\title{Solution of the Anderson impurity model via the functional renormalization group} 

\author{Simon Streib, Aldo Isidori,  and Peter Kopietz}

\affiliation{Institut f\"{u}r Theoretische Physik, Universit\"{a}t
  Frankfurt,  Max-von-Laue Strasse 1, 60438 Frankfurt, Germany}

\date{November 15, 2012}

\begin{abstract}

We show that the functional renormalization group is a numerically cheap method to obtain the low-energy behavior of the Anderson impurity model describing a localized impurity level coupled to a bath of conduction electrons. Our approach uses an external magnetic field as flow parameter, partial bosonization of the transverse spin fluctuations, and frequency-independent interaction vertices determined by Ward identities. The magnetic field serves also as a regulator for the bosonized spin fluctuations, which are suppressed at large field. We calculate the quasiparticle residue and spin susceptibility in the particle-hole symmetric case and obtain excellent agreement with the Bethe ansatz results for arbitrary coupling.

\end{abstract}

\pacs{72.15.Qm, 71.27.+q, 71.10.Pm}

\maketitle

The Anderson impurity model (AIM)
describes a localized impurity in contact with a
bath of non-interacting electrons.\cite{Anderson61}
The model was first introduced in the context of material science 
for describing the emergence of local moments in
metals and has been studied for half a century
by various methods.\cite{Hewson93}
In the past decade, renewed attention 
has been drawn to the
AIM because of its 
experimental realization in quantum dot systems. 
Moreover, the solution of the AIM is one of the
fundamental steps in the so-called dynamical mean-field theory.\cite{Georges96}
In practice, this step is often implemented  using
Wilson's numerical renormalization group (NRG), which 
yields numerically controlled results for the thermodynamic and 
spectral properties.\cite{Wilson75} 
In the 1980s the thermodynamics of the AIM has also been obtained
exactly via the Bethe ansatz (BA).\cite{Tsvelick83}  
For later reference, we quote here 
the BA results for the spin and charge susceptibilities in the particle-hole symmetric case,\cite{Zlatic83}
 \begin{subequations}
 \begin{eqnarray}
 \pi \Delta {\chi}_s & =  & \sqrt{\frac{2 }{ \pi u} }     e^{ \pi^2 u /8}
 \int_{0}^{\infty} dx e^{ - x^2/(2u)} \frac{ \cos ( \pi x /2 )}{1-x^2},
 \label{eq:bethechis}
 \\
  \pi \Delta {\chi}_c & = &\sqrt{ \frac{2}{\pi u}}    e^{ -\pi^2 u /8} 
 \int_{0}^{\infty} dx e^{ - x^2/(2u)} \frac{ \cosh ( \pi x /2 )}{1+x^2}.
 \label{eq:bethechic}
 \hspace{5mm}
 \end{eqnarray}
 \end{subequations}
Here $u = U / ( \pi \Delta )$, where $U$ is the interaction between two impurity electrons
with opposite spin,
and $\Delta$ is the hybridization energy 
to the conduction bath, in the limit where the latter has infinite bandwidth and constant density of
states. For $ u \gtrsim 2$ the charge susceptibility ${\chi}_c$
becomes exponentially small, while the spin susceptibility
${\chi}_s = (2 T_K)^{-1} $
is proportional to the inverse
Kondo temperature 
$T_K = \Delta  ( \pi u /2  )^{1/2} e^{ -\pi^2 u /8 + 1/ (2u)}$.

Although Eqs.~(\ref{eq:bethechis}, \ref{eq:bethechic})
can be confirmed numerically using  the NRG,  
it would be useful to have a numerically cheap 
method to obtain the correct low-energy physics of
the AIM at strong coupling, 
which is dominated by the exponentially small
Kondo temperature. In this work we show that
this can be achieved by means of 
the functional renormalization group (FRG) method.\cite{Pawlowski07,Kopietz10,Metzner12}
In the past few years, many authors have studied the AIM
using different versions
of the FRG,\cite{Hedden04,Karrasch06,Karrasch08,Bartosch09,Isidori10,Jakobs10,Freire12,Kinza12} 
but failed  in  reproducing  
the correct strong coupling behavior of the AIM. 
We show that this problem can be solved using a
simple truncation of the FRG hierarchy involving only
frequency-independent interaction vertices which are fixed by Ward identities (WI), 
provided that the transverse spin fluctuations are properly bosonized
and that the corresponding bosonic self-energy, obtained from a skeleton equation, fulfills the WI.

For simplicity, we focus on the particle-hole symmetric AIM.
After integrating out the conduction electrons, 
we obtain  the following Euclidean action for the Grassmann field
$d_{\sigma} ( \tau  )$ describing
the localized electron,
 \begin{equation} 
 S = - \int_{\omega} \sum_{\sigma} G_{ 0 , \sigma}^{-1} ( i \omega ) 
 \bar{d}_{\omega \sigma} d_{\omega \sigma} + U \int_0^{1/T} \!\! d \tau \, n_{\uparrow} 
 n_{\downarrow} .
 \label{eq:action}
 \end{equation} 
Here $T$ is the temperature 
(we set $k_B = \hbar =1$ and focus on the zero temperature limit below),
$\int_{\omega}  = T \sum_{\omega}$ denotes summation over fermionic
Matsubara frequencies~$ i \omega$, 
$n_{\sigma}  = \bar{d}_{\sigma} ( \tau ) d_{\sigma} ( \tau )$
represents the number of localized electrons with spin $\sigma = \uparrow, \downarrow$,
and $d_{\omega \sigma} =  \int_0^{1/T} \! d \tau \, e^{ i \omega \tau } d_{\sigma} ( \tau )$.
In the particle-hole symmetric case
the
inverse bare Green function is
$G_{0, \sigma}^{-1} ( i \omega )
  = i \omega + U/2 + \sigma H + i \Delta \mathop{\rm sgn}  \omega $,
where 
$H$ is the external magnetic field in units of energy.

Our strategy is inspired by the 
recent work by Edwards and Hewson,\cite{Edwards11}
who developed a 
field-theoretical renormalization group 
approach
which yields the correct low-energy physics
of the AIM for arbitrary coupling.
They noticed
that for large magnetic fields 
the AIM can be studied perturbatively, even for $U \gg \Delta$, 
as long as $U/H$ is sufficiently small. Following Ref.~\onlinecite{Edwards11}, let us therefore use
the external magnetic field $H$ as a flow parameter for the FRG. 
Unfortunately, using this cutoff scheme 
we could  not  reproduce the strong-coupling physics of the
AIM if we formulated the FRG exclusively in terms of 
fermionic degrees of freedom.  
The problem is that the resulting fermionic vertices 
exhibit a rather strong
frequency dependence which cannot be neglected.
To avoid this problem, we partially bosonize the action
(\ref{eq:action}) by decoupling  the interaction in terms of
a complex boson field $\chi$ representing transverse spin fluctuations.
It turns out to be advantageous, however, to bosonize only a part of the interaction, while
retaining a fermionic parametrization for the rest.
The reason is that 
the partial bosonization removes the strong frequency dependence of the interaction vertices 
only in one particular scattering channel, while the other channels can potentially exhibit 
strongly frequency-dependent vertices.
In order to avoid this phenomenon, one has to find a compromise and retain a part of the interaction 
in the original fermionic parametrization.
Therefore, we write the interaction in Eq.~(\ref{eq:action}) as
$U n_{\uparrow} n_{\downarrow} = U_{\parallel} n_{\uparrow} n_{\downarrow}
 - U_{\bot} s_+ s_-$, where $U = U_{\parallel} + U_{\bot}$,     
$s_{+}   = \bar{d}_{\uparrow}  d_{\downarrow}$, and
$s_{-}   = \bar{d}_{\downarrow}  d_{\uparrow}$.
Only the transverse part  $- U_{\bot} s_+ s_-$   is then bosonized
by introducing a complex
field $\chi$ via a usual Hubbard-Stratonovich transformation.\cite{Bartosch09,Isidori10} 
We choose $U_{\bot}$ such that it vanishes for $H \rightarrow \infty$, 
and that it reduces
to the bare interaction $U$ for $H \rightarrow 0$. Therefore we set
$U_{\bot}^{-1} = U^{-1} + R_H$ and require 
$R_{ 0 } =0$ and $R_{\infty} = \infty$. Note that
$R_H$ plays the role of a regulator
for the bosonic sector of the theory which switches on 
the interaction mediated by transverse spin fluctuations.
The proper choice of $R_H$
will be discussed in more detail below.
In comparison to the 
renormalization group method developed in Ref.~\onlinecite{Edwards11},
our FRG approach does not require the introduction of {\it ad hoc} counter terms. 

\begin{figure}[tb]    
  \centering
  \includegraphics[keepaspectratio,width=\linewidth]{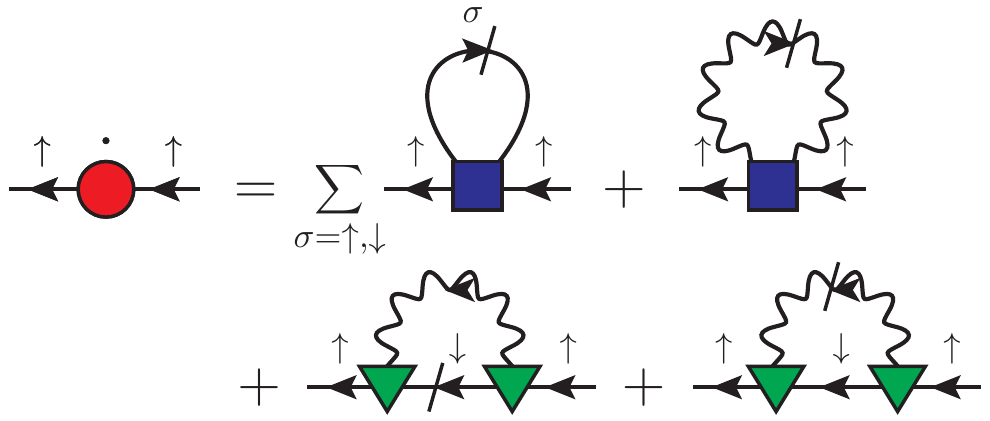}
  \caption{\label{fig:1}(Color online) %
FRG flow equation (\ref{eq:flowsigma}) for $\Sigma_{\uparrow} ( i \omega )$
(shaded circle). Solid and wavy lines represent full fermion and boson propagators, respectively;
slash insertions denote the corresponding 
single-scale propagators in Eq.~(\ref{eq:single-scale-prop}). 
The dot on the left represents a derivative with respect to $H$.
} 
\end{figure}

It is now straightforward to write down formally exact
FRG flow equations for the
irreducible vertices of our boson-fermion model.\cite{Kopietz10,Bartosch09}
For our purpose, it is sufficient to 
neglect  the frequency dependence of all vertices
with more than two external legs.  We also neglect the contribution from charge- and 
longitudinal spin fluctuations.
The resulting flow equation for the fermionic
self-energy $\Sigma_{\sigma} ( i \omega )$
is shown graphically
in Fig.~\ref{fig:1}; the corresponding analytic expression reads
 \begin{eqnarray}
 & & \frac{\partial \Sigma_{\sigma} ( i \omega ) }{\partial H }  =  
 \int_{\omega^{\prime}} \sum_{\sigma^{\prime}}
 \dot{G}_{\sigma^{\prime}} ( i \omega^{\prime} )
 {\Gamma}_{4}^{ \bar{d}_{\sigma} \bar{d}_{\sigma^{\prime}} d_{\sigma^{\prime}} d_{\sigma} }
 \nonumber
 \\
 & + &
  \int_{\bar{\omega}} \dot{F}_{\bot} ( i \bar{\omega} )
\Gamma_4^{ \bar{d}_{\sigma} d_{\sigma} \bar{\chi} \chi } 
+ \int_{\omega^{\prime}}
 \Bigl[
 \dot{G}_{- \sigma } ( i \omega^{\prime} ) 
 F_{\bot} ( i \sigma ( \omega -  \omega^{\prime}) )
 \nonumber
 \\
 &  & \hspace{5mm}
 + \,
  {G}_{- \sigma } ( i \omega^{\prime} ) 
 \dot{F}_{\bot} ( i \sigma ( \omega -  \omega^{\prime}) )
 \Bigr] 
\Gamma_3^{ \bar{d}_{\uparrow}  d_{  \downarrow} \chi  } 
\Gamma_3^{ \bar{d}_{\downarrow}  d_{  \uparrow} \bar{\chi}  } .
 \label{eq:flowsigma}
 \end{eqnarray}
The single-scale propagators in our cutoff scheme are
 \begin{eqnarray}
 \dot{G}_{\sigma} ( i \omega ) & = & - \sigma G_{\sigma}^2 ( i \omega ),
\; \;
 \dot{F}_{\bot} ( i \bar{\omega} )  =  - 
 F_{\bot}^2 ( i \bar{\omega} )\partial R_H / \partial H  , \hspace{3mm}
 \label{eq:single-scale-prop}
 \end{eqnarray}
where $G_{\sigma} ( i \omega )$ is the exact fermionic Green function 
and $F_{\bot} ( i \bar{\omega} )$ is the exact 
propagator of the spin fluctuation field $\chi$.
These propagators can be expressed in terms of the corresponding
self-energies $\Sigma_{\sigma} ( i \omega )$ and
$\Pi_{\bot} ( i \bar{\omega} )$ via the Dyson equations
$G_{\sigma}^{-1} ( i \omega ) = G_{0,\sigma}^{-1} ( i \omega ) - \Sigma_{\sigma} ( i \omega )$ and
$F_{\bot}^{-1} ( i \bar{\omega} ) = U_{\bot}^{-1} - \Pi_{\bot} ( i \bar{\omega} )$.
We denote bosonic Matsubara frequencies by $i \bar{\omega}$.
The right-hand side of 
Eq.~(\ref{eq:flowsigma}) depends on the
irreducible four-point vertices, 
${\Gamma}_4^{ \bar{d}_{\sigma} \bar{d}_{\sigma^{\prime}} d_{\sigma^{\prime}} d_{\sigma} }$ and
$\Gamma_4^{ \bar{d}_{\sigma} d_{\sigma} \bar{\chi} \chi }$, and on the three-point vertices
$\Gamma_3^{ \bar{d}_{\uparrow}  d_{  \downarrow} \chi  }=
\Gamma_3^{ \bar{d}_{\downarrow}  d_{  \uparrow} \bar{\chi}  } \equiv \gamma$.
The superscripts indicate the types of legs attached to the vertices.

Our goal is to derive flow equations for 
the quasiparticle residue $Z$ and
the spin-dependent part $M$ of the 
fermionic self-energy, defined via the low-energy expansion
 \begin{equation}
 \Sigma_{\sigma} ( i \omega ) = U/2 - \sigma M + ( 1- Z^{-1} ) i \omega
 + {\cal{O}} ( \omega^2 ).
 \label{eq:lowenergy}
 \end{equation}
In order to obtain $M$ and $Z$ from the solution
of the flow equation (\ref{eq:flowsigma})
we need to know the bosonic self-energy $\Pi_{\bot} ( i \bar{\omega} )$
and  three different vertices:
 ${\Gamma}_4^{ \bar{d}_{\sigma} \bar{d}_{\sigma^{\prime}} d_{\sigma^{\prime}} d_{\sigma} }$,
$\Gamma_4^{ \bar{d}_{\sigma} d_{\sigma} \bar{\chi} \chi } $, and $\gamma$.
One could write down additional flow equations for these vertices,
but these depend on  higher order vertices
and it is not clear how to truncate this infinite hierarchy. 
We shall therefore adopt a different strategy 
which uses WI and skeleton equations to obtain a closed
system of flow equations for $M$ and $Z$. 

First of all, we note that
the three-point vertex
can be related to $M$ via the WI
 \begin{equation} 
 \Gamma_3^{ \bar{d}_{\uparrow}  d_{  \downarrow} \chi  } =
  \Gamma_3^{ \bar{d}_{\downarrow}  d_{  \uparrow} \bar{\chi}  } \equiv 
\gamma =
 b /( h + u_{\bot} \arctan b ), 
 \label{eq:WardKoyama}
 \end{equation}
which can be derived  following
the work by Koyama and Tachiki.\cite{Koyama84,Streib13} 
Here $u_{\bot} = U_{\bot} / ( \pi \Delta )$
and  $b =  h + m$, with $m = M/\Delta$ and
$h = H / \Delta$.

Next, let us derive an alternative flow equation for
$M$ which does not explicitly involve the 
four-point vertices.
To this end we consider the 
longitudinal spin-susceptibility $\chi_{\parallel}
 = \partial s / \partial H$, where $s = \langle n_{\uparrow} - n_{\downarrow} \rangle /2$
is the local magnetic moment, normalized such that $s=1/2$ corresponds to a fully
magnetized state. The local moment is then connected to $M$
via the Friedel sum rule,\cite{Hewson93} 
 \begin{equation} 
 \pi s = \arctan ( h + m ) = \arctan b.
 \label{eq:Friedel}
 \end{equation}
Taking the derivative 
with respect to $h$ we obtain
 \begin{equation}
 \partial m / \partial h =  \chi_{\parallel} / \rho -1 ,
 \label{eq:mflow}
 \end{equation}
where $\rho = [ \pi \Delta ( 1 + b^2 ) ]^{-1}$ is the exact density of states
at vanishing energy.
To determine $\chi_{\parallel}$ we use the WI\cite{Hewson93,Kopietz10a} 
 \begin{equation} 
  \chi_{\parallel} / \rho = 1/Z + \rho \Gamma_{\bot},
 \end{equation}
where $\Gamma_{\bot}$ is the exact interaction vertex between
two electrons with opposite spin at vanishing frequencies.
In our partially bosonized theory we have
 \begin{equation}
  \Gamma_{\bot} =  \gamma^2 U_{\bot} /[1 - U_{\bot} \Pi_{\bot} (0) ]
 + {\Gamma}_4^{ \bar{d}_{\uparrow} \bar{d}_{\downarrow} 
 d_{\downarrow}  d_{\uparrow}} .
 \end{equation}
But the three-point vertex  and
the fermionic four-point vertex turn out to be mutually related via
a skeleton equation.
With the help of standard functional methods\cite{Kopietz10}
it is straightforward  to show that, at finite frequencies,
\begin{eqnarray}
 \Gamma_3^{ \bar{d}_{\uparrow} d_{\downarrow} \chi }
 ( \omega + \bar{\omega} ,  \omega ; \bar{\omega} ) & = & 1 
 - \int_{\omega^{\prime}} G_{\downarrow} ( i \omega^{\prime} )
 G_{\uparrow} ( i \omega^{\prime} + i \bar{\omega} )
  \nonumber
 \\[-1mm]
 & & \hspace{-15mm} \times 
 {\Gamma}_4^{ \bar{d}_{\uparrow} \bar{d}_{\downarrow} d_{\downarrow} d_{\uparrow}} ( \omega + \bar{\omega} , \omega^{\prime} ; \omega , \omega^{\prime} +  \bar{\omega } ).
 \label{eq:skeleton1}
\end{eqnarray}  
Using the above relation to express
${\Gamma}_4^{ \bar{d}_{\uparrow} \bar{d}_{\downarrow} 
 d_{\downarrow}  d_{\uparrow}}$ in terms of 
$ \Gamma_3^{ \bar{d}_{\uparrow}  d_{  \downarrow} \chi  } \equiv \gamma $,  
we find  
for the longitudinal susceptibility\cite{footnoteskeleton}
\begin{equation}
\frac{ {\chi}_{\parallel}}{\rho} = 
\frac{1}{Z} + \frac{1}{1+b^2} \frac{b}{\arctan b} \left[\frac{b}{h} - \frac{1}{Z} \right],
 \label{eq:chilon}
\end{equation}
where we have used the WI (\ref{eq:WardKoyama}) 
to eliminate the vertex $\gamma$.
Substituting the expression (\ref{eq:chilon})
into Eq.~(\ref{eq:mflow}),
we obtain a flow equation for $m$ which depends only on the self-energy parameters
$m$ and $Z$.

Finally, to close our system of flow equations we
approximate  the bosonic self-energy $\Pi_{\bot} ( i \bar{\omega} )$, 
appearing implicitly in Eq.~(\ref{eq:flowsigma}), by
 \begin{equation}
 \pi \Delta\Pi_{\bot} ( i \bar{\omega} ) =   \gamma  
P_{\bot} (
  | \bar{\omega} | / ( Z \Delta ) , 
  b   \mathop{\rm sgn} \bar{\omega} ),
 \label{eq:Pibotres}
 \end{equation}
where the dimensionless function
 \begin{equation}
   P_{\bot} ( x , b ) = \frac{ \ln [ 1 + x /(1-ib) ]}{ ( x - 2 i b )( 1 + x/2 - i b) }
  -  \frac{ 2i \arctan b}{x-2 i b }
 \end{equation}
is obtained from the skeleton equation\cite{Bartosch09} for 
$\Pi_{\bot} ( i \bar{\omega} )$.
The prefactor in Eq.~(\ref{eq:Pibotres}) is determined by demanding 
that $\Pi_{\bot} (0)$ satisfy the
WI\cite{Koyama84,Edwards11} 
 \begin{equation}
 \Pi_{\bot} (0) /[ 1 - U_{\bot}
 \Pi_{\bot} (0) ] \equiv \chi_{\bot} =  s / H
 \label{eq:chibot}
 \end{equation}
relating the transverse  spin-susceptibility $\chi_{\bot}$ to the local moment $s$.
Using the Friedel sum rule (\ref{eq:Friedel}) it is easy to see that our 
expression
(\ref{eq:chilon}) for the longitudinal susceptibility is
consistent with Eq.~(\ref{eq:chibot}) for 
$\chi_{\bot}$, in the sense that for  $H \rightarrow 0$
both susceptibilities approach the same limit $\chi_s$. 
We point out that Eqs.~(\ref{eq:WardKoyama})
and (\ref{eq:chibot}) imply that
$ a \equiv [ 1 - U_{\bot} \Pi_{\bot} ( 0 ) ] = h / ( h + u_{\bot} \arctan b ) > 0$,
which guarantees that there is no magnetic instability.

We have now obtained a closed system of flow equations for
$m$ and $Z$. For simplicity, we also expand the
function $P_{\bot} ( x , b )$ in Eq.~(\ref{eq:Pibotres}) to linear order
in $x$, which is consistent
with the low-energy expansion (\ref{eq:lowenergy}) 
in the fermionic sector.
After straightforward algebra we arrive at the flow equation
$h \partial  Z/ \partial h  =  \eta Z$ for the quasiparticle residue $Z$, with
 \begin{eqnarray}
 \eta & = &  h Z \gamma^2 u_\bot {\rm Im} 
 \int_0^{\infty} \frac{dx}{( 1 - i b + x )^2} \frac{c}{ (a + c x )^2 }
 \nonumber
 \\
 & + & h Z \gamma^2 u_\bot^2  \frac{\partial r }{ \partial h} 
{\rm Re} 
 \int_0^{\infty} \frac{dx}{( 1 - i b + x )} \frac{2  c}{ (a + c x )^3 }.
 \label{eq:etares}
 \end{eqnarray}
Here $r  = \pi \Delta R_H$ is the dimensionless bosonic regulator,
and
 \begin{equation}
c  =  (u_\bot \gamma)(2 i b)^{-1} \left[
 (1-ib)^{-2} - \arctan b/ b \right].
 \end{equation}

\begin{figure}[tb]    
  \centering
  \includegraphics[keepaspectratio,width=\linewidth]{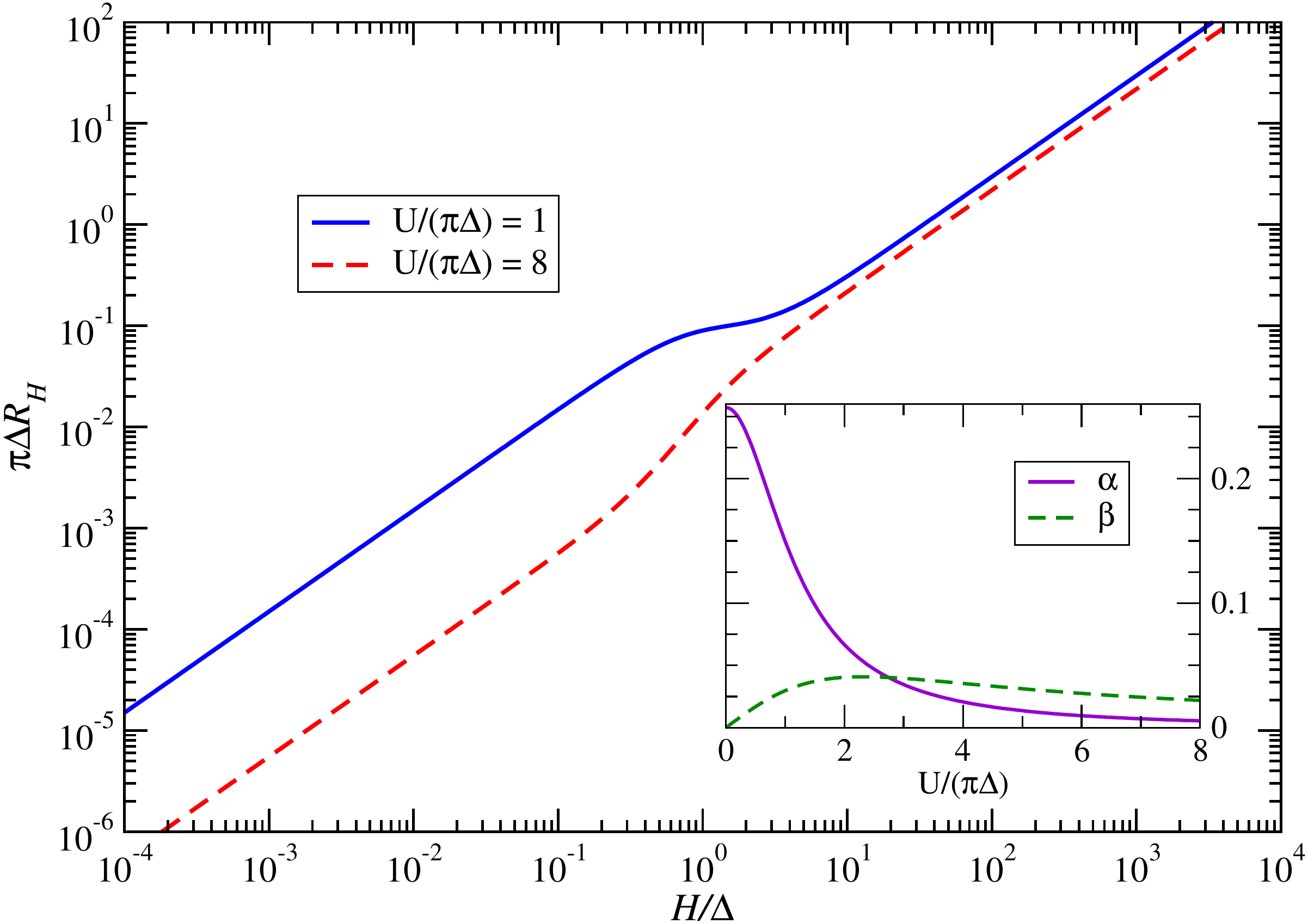}
  \caption{\label{fig:2}(Color online) %
Bosonic regulator $r=\pi \Delta R_H$
as a function of the magnetic field $h=H/\Delta$, for 
$U/(\pi\Delta) = 1$ (solid line) and $U/(\pi\Delta) = 8$ (dashed line).
The regulator is linear in $h$ at small and large $h$,
with a change in slope from $\alpha$ ($h\lesssim 1$) to $\beta$
($h\gtrsim 1$). The inset shows the $U$-dependence of $\alpha$ 
and $\beta$.} 
\end{figure}

The above system of flow equations 
can be easily solved numerically once
the regulator $ r  =  \pi \Delta R_H$ has been specified.
The simplest choice of a linear magnetic field
dependence of the bosonic regulator ($r \propto h$, in analogy with the 
regulator in the fermionic sector) does not lead to satisfactory results.
Instead, 
we found that the best optimization is obtained
with a regulator of the form
 \begin{equation}
 r = ( \alpha - \beta ) h / (1 + h^2 ) + \beta h,
 \label{eq:bosonic_cutoff}
 \end{equation}
which is linear in $h$ both for small and large $h$,
but exhibits a change in slope from $\alpha$ to $\beta$ at $h \approx 1$,
as shown in Fig.~\ref{fig:2}.
The specific form of our regulator, in Eq.~(\ref{eq:bosonic_cutoff}), 
is of course only one possible choice. 
Any other functional form which switches between two slopes as a function of the 
magnetic field yields practically identical results: 
We have checked this explicitly using the alternative function 
$r = (\alpha - \beta)\arctan h + \beta h$.
Comparing our FRG results at weak coupling, $U \ll \Delta$,
with the known perturbative expansion for $Z$ and $\chi_s$,
we also realized that the coefficients $\alpha$ and $\beta$ are weakly dependent 
functions of the dimensionless bare interaction $u = U /( \pi \Delta)$.
Their functional dependence is well described by 
rational functions of low degree. Hence, we introduce the 
following ansatz,
 \begin{equation}
 \alpha = \alpha_0/( \alpha_2 + u^2 ), \; \; \;
 \beta = \beta_1 u /( \beta_2 + u^2 ),
 \label{eq:alpha-beta}
 \end{equation}
where the four parameters $\alpha_0$, $\alpha_2$, $\beta_1$, and $\beta_2$
can be determined by matching the FRG results for $Z$ and $\chi_s$ 
at weak coupling (e.g., for $u = 0.1$) with  
the corresponding values obtained in perturbation  theory  at ${\cal O}(u^2)$, 
and by imposing that the Wilson ratio $R$
is equal to 2 for two values of $u$
in the Kondo regime (e.g., we impose $R=2$ for $u = 4$ and $u = 8$).
Note that in the AIM the expression for the Wilson ratio reads
\begin{equation}
R \equiv \frac{{\chi}_s(U)/{\chi}_s(0)}{Z^{-1}(U)/Z^{-1}(0)} 
  = \pi \Delta {\chi}_s Z = \frac{2{\chi}_s}{{\chi}_s + {\chi}_c},
\end{equation}
where the last equality follows from the Yamada-Yosida
WI.\cite{Kopietz10a,Yamada75}
At large $U$, charge
fluctuations are 
strongly suppressed and
the low-energy physics of the AIM is effectively the same as in the Kondo model. 
In this regime the charge susceptibility ${\chi}_c$ is then
negligible and the Wilson ratio takes the Kondo-model value $R=2$. 
For the parametrization of $\alpha$ and $\beta$ in
Eq.~(\ref{eq:alpha-beta}), 
we find $\alpha_0 \approx 0.36$,
$\alpha_2  \approx 1.4$, $\beta_1 \approx 0.19$, and $\beta_2 \approx 5.4$.
The $U$-dependence of $\alpha$ and $\beta$ is shown in the inset of
Fig.~\ref{fig:2}.
As a final remark, we note that
in principle it should be possible to fix the interaction dependence of $\alpha$ and $\beta$ 
entirely from a fit to the weak coupling expansion, 
but we found that this fit is numerically unstable. 
Instead, the requirement that at strong coupling the Wilson ratio 
equals 2 for two (arbitrary) values of the interaction 
yields numerically stable constraints on the regulator. 
More complicated regulators, involving more parameters, are of course possible, 
but our four-parameter regulator defined in Eqs.~(\ref{eq:bosonic_cutoff}) and (\ref{eq:alpha-beta}) 
is the simplest one that is able to interpolate between 
some known weak- and strong-coupling behaviors: 
The perturbative weak-coupling expansion, and the asymptotic value of the Wilson ratio, 
which are both a priori accessible even in more complex models 
for which no exact solution is available.

\begin{figure}[tb]    
  \centering
  \includegraphics[keepaspectratio,width=\linewidth]{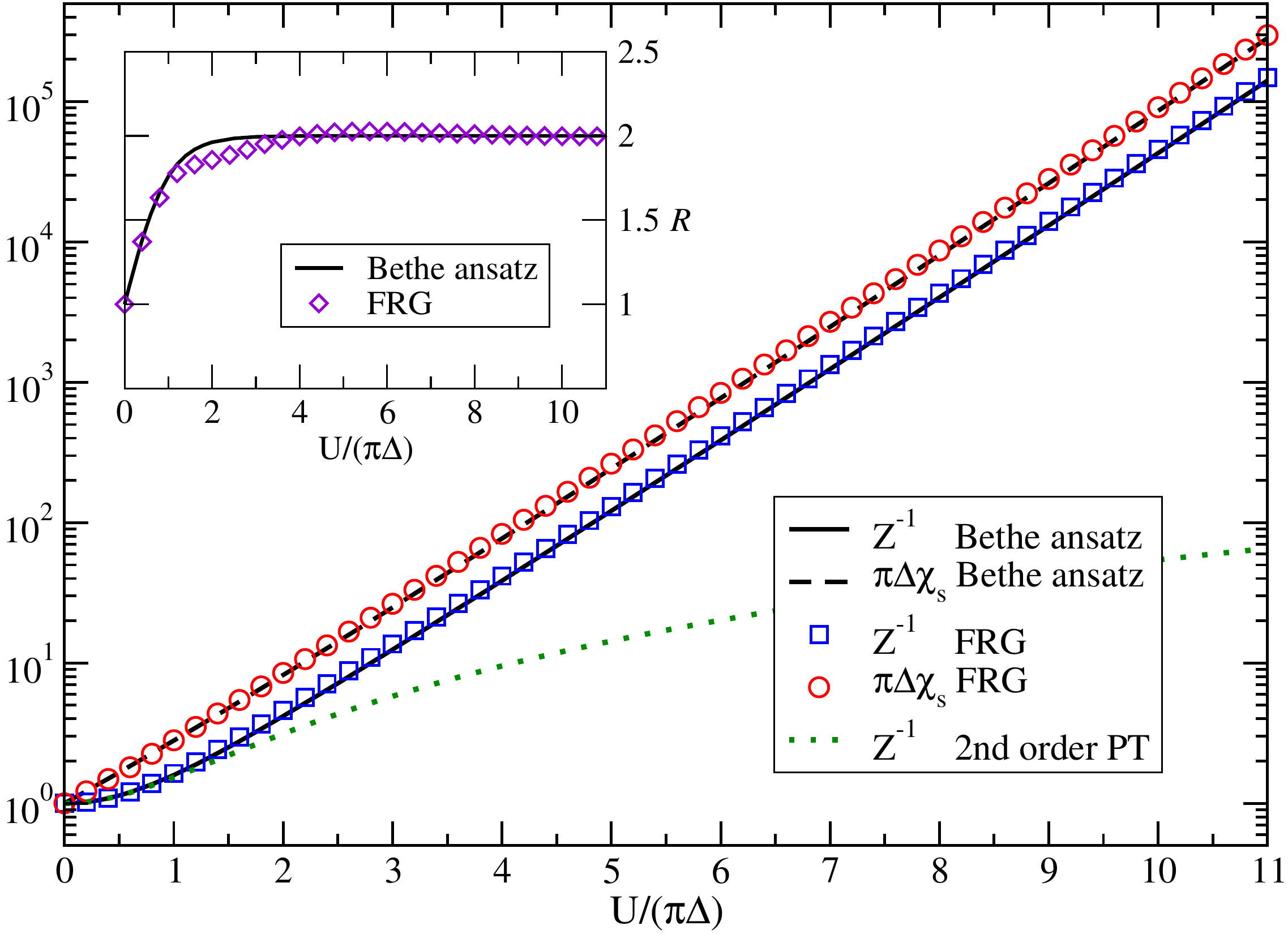}
  \caption{\label{fig:3}(Color online) %
Quasiparticle residue $Z^{-1}$ and spin susceptibility $\chi_s$
as functions of $U/(\pi\Delta)$ for $H=0$. The FRG results for $Z^{-1}$ (squares) and $\chi_s$ (circles)
are compared with the corresponding BA results (solid and dashed lines, respectively).
The dotted line represents $Z^{-1}$ within second-order perturbation theory. 
The inset shows the FRG (diamonds) and BA (solid line) results for the Wilson ratio $R$.
} \end{figure}

Having fixed the bosonic regulator, 
we can now solve the flow equations numerically,
with minimal computational cost.
Choosing the initial magnetic field $H_0$
sufficiently large, we may use
as initial conditions for $M$ and $Z$ the
Hartree-Fock values $M_0=U/2$ and $Z_0=1$, which are exact in the limit $H_0\to\infty$.
The FRG results for $Z$ and $\chi_s$, at $H=0$, are 
shown in Fig.~\ref{fig:3}.
The remarkable agreement of our approach with the BA results
shows that the present FRG
truncation, supplied with exact WI, is indeed able to capture  
the exponentially small Kondo scale.

\begin{figure}[tb]    
  \centering
  \includegraphics[keepaspectratio,width=\linewidth]{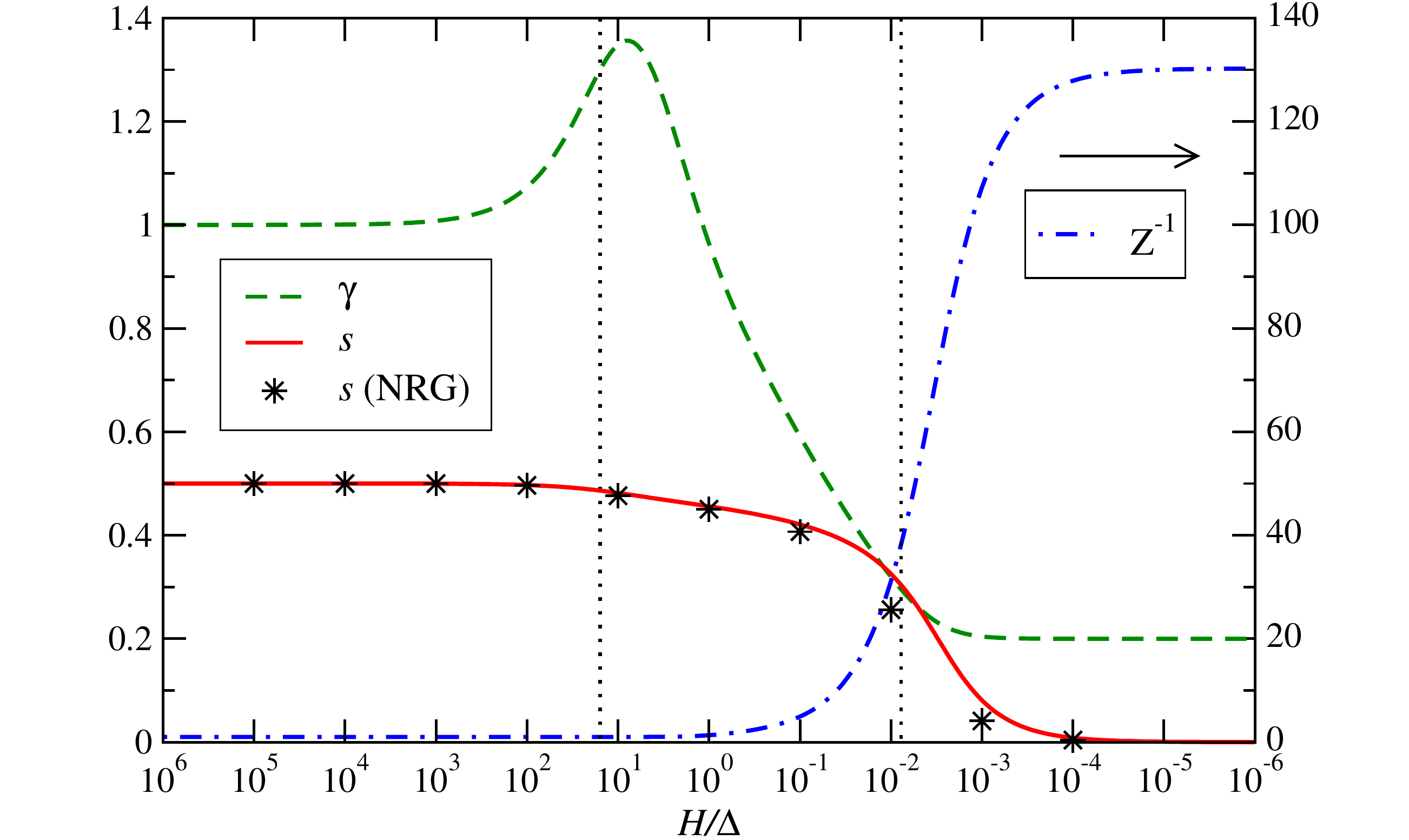}
  \caption{\label{fig:4}(Color online) %
Quasiparticle residue $Z^{-1}$ (dot-dashed line, right scale), 
magnetic moment $s$ (solid line),
and three-point vertex $\gamma$ (dashed line)
as functions of $H$ 
for $U/(\pi\Delta)=5$, calculated with FRG. 
The vertical dotted lines indicate the crossover scales 
$H=U$ and $H=Z\Delta\approx T_K$.
Symbols denote the corresponding NRG results for $s$.
} 
\end{figure}

In Fig.~\ref{fig:4} we show, 
for $u=U/(\pi\Delta)=5$,
the magnetic field dependence of
$Z$, $s$, and $\gamma$.  
In the magnetization curve $s(H)$ one can clearly identify 
the two energy scales characterizing the strong coupling regime of the AIM,
namely the bare interaction $U$ and the width $Z\Delta \approx T_K$ of the Kondo resonance.
Indeed, reducing the external magnetic field $H$ from a large initial value, 
at $H \approx U$ 
we observe  a slow logarithmic decrease in the magnetic moment
from the initial value $s = 1/2$.
This regime, where $ U \gtrsim H \gtrsim T_K $,
corresponds to the  localized  moment regime:\cite{Tsvelick83}
Here $s(H)$ behaves  as a rational function of $\ln(H/T_K)$.
The decrease in $s$ becomes then
faster when $H \approx Z\Delta$ (Kondo regime), until the magnetization vanishes linearly,
as $s(H) = \chi_s H$, for $H \ll Z\Delta$.
The presence of two different energy scales is also reflected in the non-monotonic behavior
of the three-point vertex $\gamma(H)$. Starting from the bare value $\gamma=1$ at large $H$,
the vertex becomes initially larger than unity for $H\approx U$ 
and later decreases again, approaching (at strong coupling) 
the value $1/u < 1$ for $H \lesssim Z\Delta$.

In summary, we have developed an FRG approach to the AIM
which uses the external magnetic field $H$ as a physical flow parameter, 
and where the fermionic interaction $U$ 
is partially decoupled via a bosonic field describing 
transverse spin fluctuations. 
The latter are controlled by an $H$-dependent regulator,
which suppresses the transverse spin fluctuations at large $H$.
We have truncated the 
FRG flow equations
keeping only frequency-independent vertices
and expanding the fermionic and bosonic self-energies
to linear order in frequency.
With the help of 
WI we have
expressed all the relevant vertices in terms of self-energy parameters, 
thereby avoiding
further approximations in the flow equations for the vertices.
Comparing our results with the
BA solution,
we have shown
that the present approach is 
able to reproduce the 
exponential $U$-dependence of the Kondo scale.
Moreover, the use of the magnetic field as a 
flow parameter gives access to the $H$-dependence of physical observables
such as the local magnetization.
Possible extensions of the present method include finite temperature calculations, the study
of the non-symmetric AIM, and 
a generalization of our approach to non-equilibrium. 
In each of these cases,
one needs to generalize the WI used in this paper
to the relevant situation. At finite temperatures, and in the
non-symmetric AIM, the modified WI can be derived
using the same functional methods used here.\cite{Kopietz10}
For systems out of equilibrium, instead, the work by Oguri\cite{Oguri01} shows that
it is possible to generalize  the WI for the AIM  to non-equilibrium
using the Keldysh diagrammatic formalism.

We thank A. C. Hewson for useful discussions and for making his unpublished
notes on the AIM available to us.
This work was 
supported by the DFG via FOR 723.


\begin{thebibliography}{99}
\bibitem{footnoteskeleton}
Since we are neglecting the frequency dependence
of all vertices,
we need to introduce a correction factor
in Eq.~(\ref{eq:skeleton1}). This can be fixed
by demanding that, for $h \rightarrow 0$, 
$\chi_{\parallel}$ in Eq.~(\ref{eq:chilon}) 
approach the same limit as $\chi_{\bot}$ 
in Eq.~(\ref{eq:chibot}).
\end{thebibliography}
\end{document}